\documentclass[prx, reprint, longbibliography, superscriptaddress, showkeys]{revtex4-2}
\usepackage{graphicx,amsfonts,amsmath,amssymb,amstext,bm,dsfont}
\usepackage{float,wrapfig}
\usepackage{subfigure,psfrag}
\usepackage{xcolor}
\usepackage[colorlinks=true,urlcolor=blue,citecolor=blue,linkcolor=blue]{hyperref}
\hypersetup{breaklinks=true}
\usepackage{tcolorbox}
\usepackage{soul}

\usepackage{caption,setspace}

\begin{document}

\title{Multiband condensate of magnons in two dimensions}
\date{August 31, 2026}

\author{Joe Bailey}
\thanks{Lead author}
\affiliation{Paul Scherrer Institute, Villigen PSI CH-5232, Switzerland}
\affiliation{Institut de Physique, \'{E}cole Polytechnique F\'{e}d\'{e}rale de Lausanne, Lausanne CH-1015, Switzerland}

\author{Pavlo Sukhachov}
\thanks{P.S. and K.B. contributed equally to this work.}
\affiliation{Nordita, Royal Institute of Technology, Stockholm University, Stockholm SE-106 91, Sweden}
\affiliation{Present address: Department of Physics and Astronomy; MU Materials Science \& Engineering Institute; University of Missouri, Columbia, MO 65211, USA}

\author{Korbinian Baumgaertl}
\thanks{P.S. and K.B. contributed equally to this work.}
\affiliation{Laboratory of Nanoscale Magnetic Materials and Magnonics, Institute of Materials, \'{E}cole Polytechnique F\'{e}d\'{e}rale de Lausanne, Lausanne 1015, Switzerland}
\affiliation{Institute of Microengineering, \'{E}cole Polytechnique F\'{e}d\'{e}rale de Lausanne, Lausanne 1015, Switzerland}

\author{Simone Finizio}
\affiliation{Paul Scherrer Institute, Villigen PSI CH-5232, Switzerland}

\author{Sebastian Wintz}
\affiliation{Paul Scherrer Institute, Villigen PSI CH-5232, Switzerland}
\affiliation{Helmholtz-Zentrum Berlin f\"{u}r Materialien und Energie, Berlin 14109, Germany}

\author{Carsten Dubs}
\affiliation{INNOVENT e.V. Technologieentwicklung; Jena, 07745 Jena, Germany}

\author{J\"org Raabe}
\affiliation{Paul Scherrer Institute, Villigen PSI CH-5232, Switzerland}

\author{Dirk Grundler}
\affiliation{Laboratory of Nanoscale Magnetic Materials and Magnonics, Institute of Materials, \'{E}cole Polytechnique F\'{e}d\'{e}rale de Lausanne, Lausanne 1015, Switzerland}
\affiliation{Institute of Microengineering, \'{E}cole Polytechnique F\'{e}d\'{e}rale de Lausanne, Lausanne 1015, Switzerland}

\author{Alexander Balatsky}
\affiliation{Nordita, Royal Institute of Technology, Stockholm University, Stockholm SE-106 91, Sweden}
\affiliation{Department of Physics, Institute for Materials Science, University of Connecticut, Storrs, CT 06269}

\author{Gabriel Aeppli}
\thanks{Lead author}
\email{aepplig@ethz.ch}
\affiliation{Paul Scherrer Institute, Villigen PSI CH-5232, Switzerland}
\affiliation{Institut de Physique, \'{E}cole Polytechnique F\'{e}d\'{e}rale de Lausanne, Lausanne CH-1015, Switzerland}
\affiliation{Department of Physics and Quantum Center, ETH Z\"{u}rich, Z\"{u}rich CH-8093, Switzerland}

\begin{abstract}
Wavelike bosonic particles can accumulate in a single mode characterized by a particular wavelength, and such condensates are at the heart of phenomena such as superconductivity and superfluidity where usually a single complex number describes their state. If there are several flavors of excitations or particles, a vector containing several complex numbers can characterize multicomponent condensation, thus opening new possibilities for textures, dynamics, and devices. Thus far, multicomponent condensates have long represented a rewarding subfield of cold atom physics, as well as a theme for research on exotic superconductors where the constituent bosons are not atoms but electron pairs. Here we consider the case where the bosons are magnons (collective spin excitations), injected by microwaves into high-quality yttrium iron garnet (YIG) crystals. Recent advances in fabrication yield thin (128 nm) films of sufficiently high quality to display multiple magnon bands quantized along the film normal. Microwave pumping can populate these bands, providing a new two-dimensional multiband condensate optimized in a narrow range of powers and frequencies due to a four-magnon scattering resonance. We establish a phase diagram for this magnonic system, reminiscent of that for exotic superconductors, revealing both single and multiband condensation.
\end{abstract}


\maketitle

\section*{Significance}
\emph{Bose-Einstein condensation, superconductivity, and even lasing can be understood as the tendency of identical particles or waves to accumulate in a single-momentum state. A natural way to relax the indistinguishability requirement is to mix two species of waves. Experiments to date have relied on ultralow temperatures. We show here that a model two-species system can be readily realized at room temperature via microwave pumping of spin waves with different thickness quantum numbers in thin crystals of a ferromagnet, yttrium iron garnet. Data and theory reveal that interactions between the two species promote multispecies condensation. The success of the experiment relies on the quality of the crystals, which, together with the results themselves, provide starting points for future microwave technology.}

\section{Introduction}

Multiband systems, where particles occupying different types of quantum states coexist and interact, are an important theme in modern many-body physics as realized for solids as well as cold atoms, providing routes both to new superfluids~\cite{Hofer-Folling-ObservationOrbitalInteractionInduced-2015} as well as high temperature and other exotic superconductivity~\cite{Binnig-Bednorz-TwoBandSuperconductivityNbDoped-1980, Rinott-Kanigel-TuningBCSBECCrossover-2017, Mukuda-E.Haller-MultibandSuperconductivityHeavy-2009, Blumberg-Karpinski-ObservationLeggettsCollective-2007, Singh-Bergeal-GapSuppressionLifshitz-2019, Mikkelsen-Flensberg-HybridizationSuperconductorSemiconductorInterfaces-2018}. They also lead to additional dynamical modes and topological phenomena, appearing because vectors rather than scalars are needed to encode the amplitudes for multiple bands. 

Condensates can occur at equilibrium, as for superconductors or superfluids at low temperatures, or after pumping populations of interacting bosons, as for polaritons in semiconductor quantum wells ~\cite{Christopoulos-Grandjean-RoomTemperaturePolaritonLasing-2007, Carusotto-Ciuti-QuantumFluidsLight-2012, Struve-Esmann-RoomtemperaturePolaritonCondensate-2026} and magnons in ferromagnets ~\cite{Demokritov-Slavin:2006}. Two-dimensional (2D) systems emerging between interfaces or at surfaces are a fertile venue for multiband condensates because confinement along the direction perpendicular to the interface results in distinct two-dimensional bands. Here we take advantage of advances in the preparation of thin films of yttrium iron garnet with ultrahigh ferromagnetic resonance quality factors to examine the multiband magnon condensation which follows from microwave pumping at room temperature.  

\begin{figure*}[t]
	\centering
	\includegraphics[width=.9\linewidth]{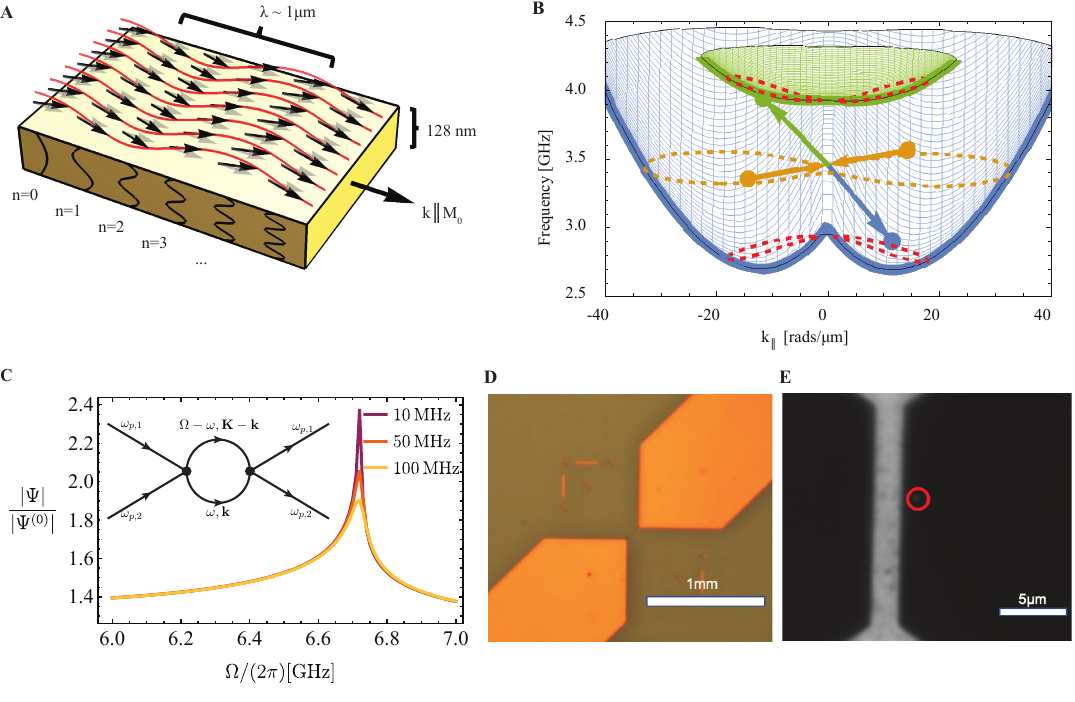}
	\caption{\textbf{Magnon dispersion and dynamics in thin yttrium iron garnet (YIG) films.} 
		\textbf{(A)} Backward volume dipole-exchange magnons ($k \parallel M_{0}$). 
		\textbf{(B)} The first two modes with an external field of $B  = 50\ mT$, see also SI Appendix. A $k \parallel M_{0}$ cut-plane (thick blue, $n = 0$, and green, $n = 1$, lines) through their bowl-like dispersions (the surface in $k_{\bot}$ is denoted by the light blue and green shadings) is shown along with the 4-magnon resonant scattering pathway. 
		\textbf{(C)} Normalized absolute value of the resummed 4-magnon scattering amplitude $\Psi_{1,2;3,4}$ for $\delta$-function broadening of 10, 50 and 100 MHz. The inset shows the Feynman diagram for the one-loop correction used in the resummation, see also SI Appendix. The amplitude is enhanced near the resonant pump-frequency condition $\Omega \approx \omega_{min\ n = 0} + \omega_{min\ n = 1}$.
		\textbf{(D)} An optical microscope overview of the device. The large light yellow features are the copper stripline contact pads, the darker region is the YIG film. 
		\textbf{(E)} shows a close-up of the stripline (narrow vertical bar) used for exciting spin waves, the red circle denotes the BLS laser spot position.
	}
	\label{fig:fig1}
\end{figure*}

\section{Concepts}

Figure~\ref{fig:fig1} illustrates the basic physics of our experiments. We start by considering magnons in a thin ferromagnetic film for which backward volume (BV) dipole-exchange magnon modes are quantized over the thickness of the film (see Fig.~\ref{fig:fig1}\textbf{A}), leaving multiple, nondegenerate two-dimensional bosonic bands. Figure~\ref{fig:fig1}\textbf{B} shows the bowl-like dispersion surfaces as a function of the in-plane wavevector $k$ for the two lowest modes (referred to by the indices, $n = 0$ and $n = 1)$. The key features are minima away from the $k = 0$ zone center, which exist for both modes although they are substantially shallower for the first excited quantum well state; the spontaneous coherent occupancy of such minima corresponds to the condensation of magnons. At relatively high microwave powers, magnons are injected via parametric pumping~\cite{Suhl:1957, Bracher-Hillebrands:2017-review}, appearing in pairs with opposite momentum, and with net energy equal to the pump frequency. Given that there are two nearby magnon bands, there is the possibility that the magnons appear in a single band (orange arrows in Fig.~\ref{fig:fig1}\textbf{B}), or in different bands (green and blue arrows in Fig.~\ref{fig:fig1}\textbf{B}). Because the $n = 1$ magnons are due to a pair of out-of-phase planar magnetization oscillations and therefore carry negligible dipole moment coupling to the microwaves, the latter process is at least an order of magnitude weaker than the former, resulting in a two orders higher threshold for pair production, as calculated in SI Appendix. The subsequent formation of condensates then depends on scattering among magnons to feed the band minima. In our two-band system, we can observe further resonant four-magnon (two magnons in, two magnons out) scattering, with the scattering amplitude between two pumped magnons into two magnons, one in each of the upper and lower bands, being maximized for pumped magnon pairs at a resonance frequency which when halved lies approximately midway between the minima of the two bands. This resonance occurs when the energy and momentum of two pumped magnons match those of an intermediate two-magnon state, i.e., when the internal propagator in the loop diagram, which quantifies the first-order correction to the four-magnon scattering amplitude, becomes nearly on-shell. Figure~\ref{fig:fig1}\textbf{C} shows the corresponding Feynman diagram and the resummed four-magnon scattering amplitude $\Psi_{1,2;3,4}$. Here,
$\Omega = \omega_{1} + \omega_{2} = \omega_{3} + \omega_{4}$ and
$\mathbf{K} = \mathbf{k}_{1} + \mathbf{k}_{2}\mathbf{=}\mathbf{k}_{3} + \mathbf{k}_{4}\mathbf{\approx 0}$ are the total frequencies and momenta of the scattered magnons created by a single photon. 
The tree-level amplitude $\Psi_{12;34}^{(0)}$ is nonresonant and varies smoothly with frequency and momentum~\cite{Lvov:book}. The pronounced frequency dependence arises from the loop contribution, whose internal two-magnon propagator becomes nearly on-shell when $\Omega \approx \omega_{min\ n = 0} + \omega_{min\ n = 1}$. We therefore calculate the loop correction and resum the corresponding loop diagrams. This can produce a pronounced enhancement of the full scattering amplitude near the two-branch resonance, as is shown in Fig.~\ref{fig:fig1}\textbf{C}, see SI Appendix for details. \\

\section{Experiments}

The material used for our experiments is YIG, a synthetic garnet with the lowest known magnon damping of any material and therefore heavily researched, including applications to magnonic and quantum devices~\cite{Demokritov-Slavin:2006, Chumak-Hillebrands-MagnonTransistorAllmagnon-2014, Forster-Wintz-DirectObservationCoherent-2019, Forster-Schutz-NanoscaleXrayImaging-2019, Wang-Chumak-MagnonicDirectionalCoupler-2020}. Perhaps most notably, in 2006 it was discovered by Demokritov et al.~\cite{Demokritov-Slavin:2006} to host a room-temperature quasi-equilibrium magnon condensate. Furthermore, due to the low spin-wave damping, nonlinear parametric processes are readily accessible in YIG~\cite{Melkov-Sholom:1989}. Recent works \cite{Schneider-Chumak-BoseEinsteinCondensation-2020, Verba-Slavin:2021} on spatial confinement in thin films, waveguides, and nanodots have revealed a complex interplay of resonances, parametric processes (both frequency degenerate and nondegenerate), four-magnon and phonon-magnon interactions which could influence condensate formation and stability. While the extent to which the room temperature condensate should be understood as a genuine, stable BEC is a matter of debate~\cite{Ruckriegel-Kopietz-RayleighJeansCondensationPumped-2015, Tupitsyn-Burin-StabilityBoseEinsteinCondensates-2008}, there are both theoretical grounds~\cite{Rezende-TheoryCoherenceBoseEinstein-2009, Troncoso-Nunez-DynamicsSpontaneousCoherence-2011, Borisenko-Demokritov-DirectEvidenceSpatial-2020} and growing evidence for BEC-related properties such as spatial coherence~\cite{Nowik-Boltyk-Demokritov:2012}, supercurrents~\cite{Bozhko-Hillebrands-SupercurrentRoomtemperatureBose-2016} and sound modes associated with the condensate order parameter~\cite{Tiberkevich-Slavin-ExcitationCoherentSecond-2019, Bozhko-Hillebrands:2019}.

We use Brillouin light scattering (BLS) to directly measure the magnon populations in thin ultraclean films of YIG with the thickness 128 nm, subjected to a static external magnetic field and excited by microwaves delivered through a micropatterned transmission line, see SI Appendix, shown in Figures~\ref{fig:fig1}\textbf{D} and \ref{fig:fig1}\textbf{E}. The distinguishing feature of the current BEC experiments is the combination of low damping (with Gilbert damping constant $\alpha = 5\times10^{-5}$) and small thickness, which allows for much narrower magnon linewidths and therefore enables the resolution of thickness-quantized magnon modes, seen previously for similarly prepared films excited by microwave bursts ~\cite{Forster-Schutz-NanoscaleXrayImaging-2019}.

\begin{figure}[t]
	\centering
	\includegraphics[width=.92\linewidth]{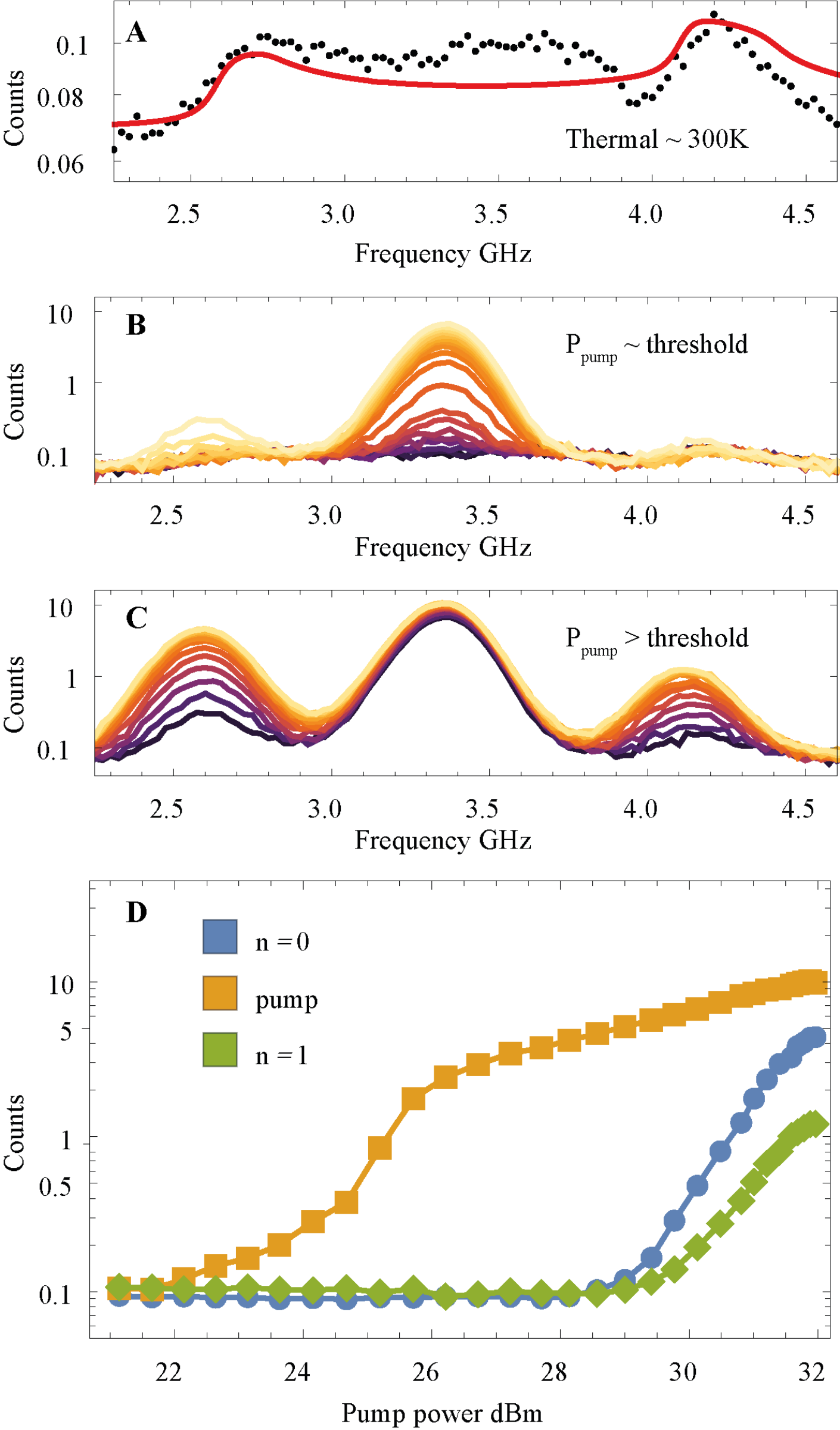}
	\caption{\textbf{Magnon occupation for different pumping regimes.}
		\textbf{(A)} Thermal spectrum of magnons in the thin YIG film, showing the broad $n = 0$ mode, and the narrower higher-frequency $n = 1$ mode, the red line is the fitted spectral function, see also SI Appendix. \textbf{(B)} As the power is increased for a fixed pump frequency of 6.7 GHz, a population of magnon pairs due to parametric pumping appears at half the pump frequency $\sim$3.5 GHz, and then the first signs of a population forming at the $n = 0$ minima $\sim$2.6 GHz follow. \textbf{(C)} As the pump power is increased further, magnon populations appear at both the $n = 0$ and $n = 1$ minima. \textbf{(D)} Signals for the different magnon populations as a function of pump power on a log-log scale, the onset of the condensation occurs $\sim$6~dBm after the pumping threshold for visible magnon pairs.}
	\label{fig:fig2}
\end{figure}

\begin{figure*}[t!]
	\centering
	\includegraphics[scale=0.35]{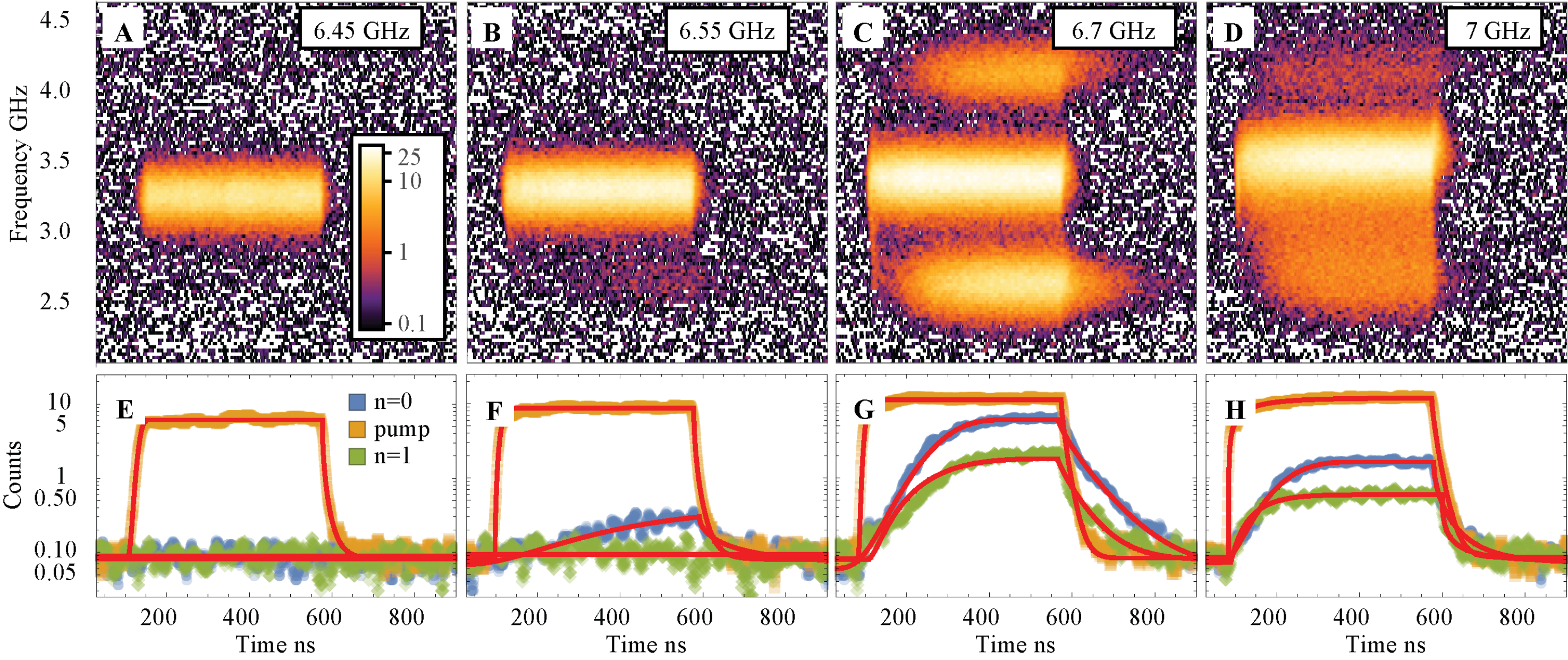}
	\caption{\textbf{Time evolution of magnon populations in quantized thickness modes.} 
		\textbf{(A)-(D)} Temporal evolution of the full magnon spectrum for different pump frequencies. \textbf{(E)-(H)} Temporal evolution at magnon frequencies corresponding to the $n = 0$ (blue) and $n = 1$ (green) minima and pumped (orange) populations. The red lines are fits of a piecewise function, see also SI Appendix.}
	\label{fig:fig3}
\end{figure*}

Figure~\ref{fig:fig2} shows spectra as a function of microwave pump power with a fixed frequency of 6.7 GHz, the resonance condition described by Fig.~\ref{fig:fig1} for our films with a static field $B = 50\ mT$. The microwave drive is a burst excitation of length 500~ns and period 2.5~$\mu s$. Without microwave excitation (Fig.~\ref{fig:fig2}\textbf{A}), we observe thermal occupation of magnons. The much broader spectrum for the $n = 0$ mode, with a cut-off at its minimum, and a sharper maximum for $n = 1$ are consistent with the dispersion surfaces of Fig.~\ref{fig:fig1}\textbf{B}, which yield the theoretical density of states given by the solid line through the data, see also SI Appendix. As the pump power is increased, we see the pair-wise formation of magnons in the lower band, at half the pump frequency (3.35~GHz), visible as the middle band/peak in Fig.~\ref{fig:fig2}\textbf{B}. Following a further rise in pump power, magnon populations appear as sharp peaks (note the logarithmic scale) in Fig.~\ref{fig:fig2}\textbf{C} at the $n = 0$ and $n = 1$ minima. The observation of this pair of peaks represents evidence for a multicomponent magnon condensate; the peaks are resolution-limited, as was the single-band peak in the original discovery of Ref.~\cite{Demokritov-Slavin:2006}. The signal corresponding to the different magnon populations plotted against the pump power reveals two thresholds at $\sim$22~dBm ($\sim$150~mW) for the parametric two-magnon pumping and $\sim$29 dBm (800~mW) for the spontaneous formations at the $n = 0$ and $n = 1$ minima. A close inspection reveals that rapid $n = 0$ population enhancement first appears at powers just below threshold for $n = 1$ enhancement (see Fig.~\ref{fig:fig2}\textbf{D}, note the log-log scale).

\begin{figure}
	\centering
	\includegraphics[width=.75\linewidth]{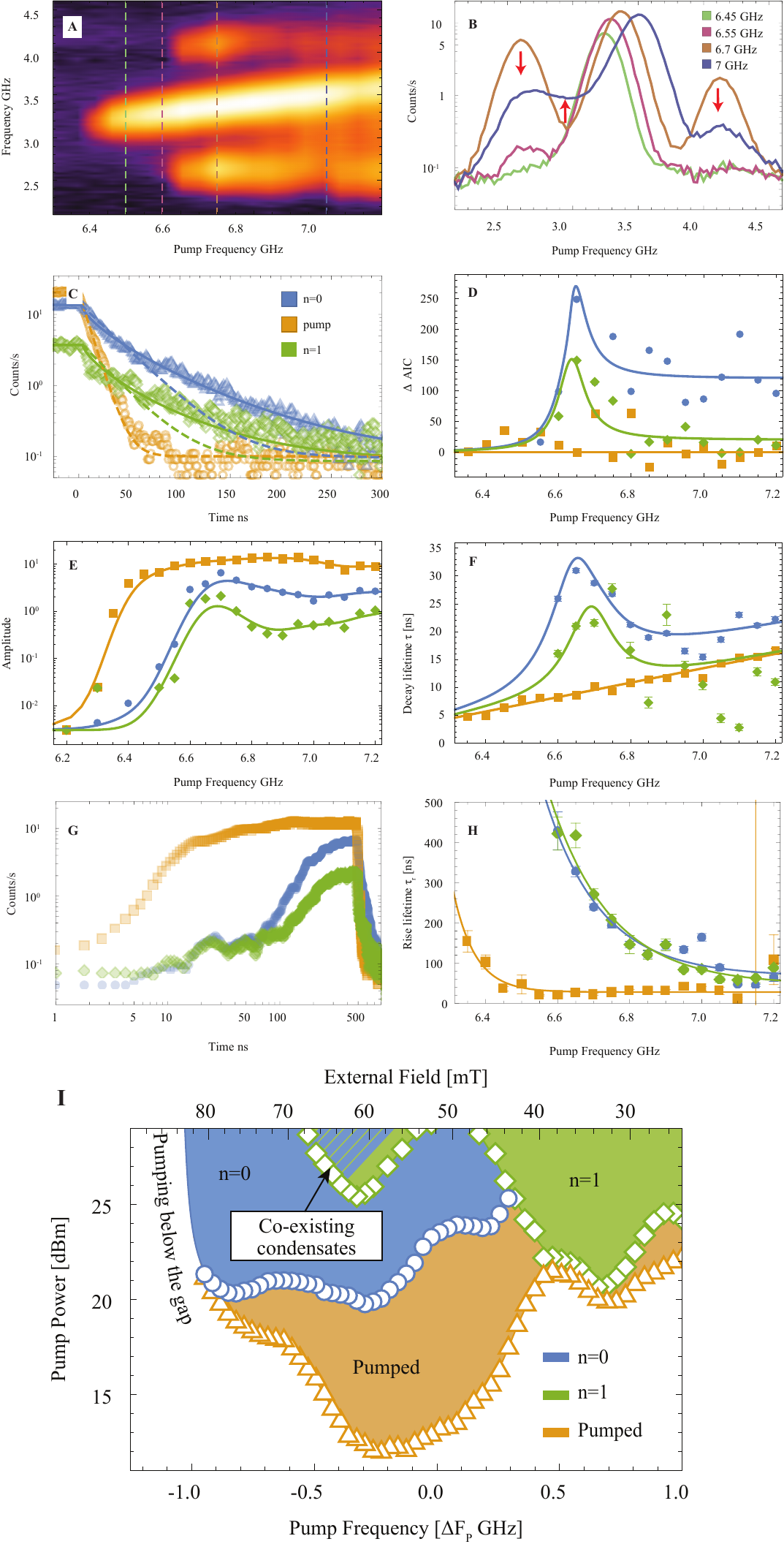}
	\caption{
		\scriptsize
		\textbf{Dependence of magnon populations on pump frequency.} 
		\textbf{(A)} Magnon spectra as a function of pump frequency integrated over time while the pump is applied. The color scale represents the logarithm of the magnon density. \textbf{(B)} Magnon spectra at fixed pump frequencies (slices denoted by corresponding dashed lines in \textbf{(A)}): 6.45~GHz where only the pumped population is excited, below resonance, where the $n = 0$ condensate is first observed (6.55~GHz), at resonance, where the multiband condensate (6.7~GHz) is formed and above resonance (7.05~GHz). For the latter, the magnons decay through conventional nonresonant scattering down to the $n = 0$ mode, as visible by the signal growth at the intermediate frequencies highlighted by the red arrows. \textbf{(C)} Decay of the different populations at resonance (time relative to the removal of the pump). While the pumped population can be fitted with a standard exponential (dashed lines -- yellow), the $n = 0$ and $n = 1$ populations require a stretched exponential defined in Eq.~\ref{eq:stretch-exp-def} (solid lines -- blue and green). \textbf{(D)} The difference in the Akaike information criterion (AIC) for models where $n_{d}$ is fixed at 1 and 0.7, showing that the decays of $n = 0$ and $n = 1\ $populations, most notably on resonance, are best described by a stretched exponential decay commonly associated with (quasi)ordered 2D systems. \textbf{(E)-(F)} Amplitudes and decay times of the different magnon populations. We observe a clear peak in both amplitude and decay lifetime of the $n = 0$ and $n = 1$ populations at the resonant pumping conditions. \textbf{(G)} Rise of the different populations at the resonance (time relative to the onset of the pump) shown on log-log scale. (\textbf{H}) The rise times from fit to different pump frequencies with piecewise equation, see SI Appendix. Solid lines are guides to the eye and error bars are statistical errors from the fits.
		\textbf{(I)} Phase diagram for multiband populations as a function of pumping power and frequency relative to the bisection of the $k = 0$ frequencies for $n = 0$ and $n = 1$ bands. The points are thresholds extracted from field and pump power sweeps for a pump frequency 7.15 GHz, see SI Appendix. In the orange region, magnon pairs generated by the nonlinear parametric pumping are apparent. For the blue and green regions, there is occupation of the $n = 0$ and $n = 1$ minima, respectively; multiband occupation characterizes the hatched region.
	}
	\label{fig:fig4}
\end{figure}

The temporal evolution of the magnon spectrum and distinct magnon populations for different pump frequencies provides further insight into the condensation of the magnons (see Fig.~\ref{fig:fig3}). For the lowest pump frequency (6.45 GHz, first column in Fig.~\ref{fig:fig3}), we only see magnons at half the pump frequency and no spontaneous populations forming elsewhere. As the pump frequency is increased to satisfy the 4-magnon resonant condition (see Fig.~\ref{fig:fig1}), we see three distinct peaks, due to the directly pumped magnons and those at the $n = 0$ and $n = 1$ minima, which have drastically enhanced decay times. For higher pump frequencies, there are decreases in the magnon occupation and decay times at both the $n = 0$ and $n = 1$ minima. Furthermore, the magnons in the $n = 0$ band yield a continuum, as commonly seen in the conventional nonresonant thermalization for single-band systems~\cite{Dzyapko-Demokritov-BoseEinsteinCondensation-2011}. In our case, this is manifested in the smearing of the three peaks as is clearly shown in Figs.~\ref{fig:fig4}\textbf{A} and \ref{fig:fig4}\textbf{B}.

We can check the conjecture of an optimized condensate at resonance by looking in more detail at the time traces. In particular, a cursory examination of the traces in Figs.~\ref{fig:fig3}\textbf{E}-\ref{fig:fig3}\textbf{F} reveals that, at resonance, these decays are not only the slowest but also that they seem nonexponential. To more precisely characterize the temporal evolution of the condensates, we fit a stretched exponential with the onset time $t_0$, the lifetime $\tau$, and the stretching exponent $n_{d}$ to the decay of different magnon populations. We use
\begin{equation}
	\label{eq:stretch-exp-def}
	f(t)= c + a e^{-\left(\dfrac{t - t_{0}}{\tau}\right)^{n_d} }.
\end{equation}
The amplitude $a$ and background $c$ are extracted directly from the data. Additional details of the fitting are provided in SI Appendix.

The stretching exponent parameterizes a range of behaviors including those typically found for single-particle excitations coupled to a bath ($n_{d} = 1$) and those related to the growth and decay of collective order in interacting many-body systems, for example 2D Ising magnets ($n_{d} = 0.5)\ $and polariton condensates ($n_{d} \approx 0.7$)~\cite{huseDynamicsDropletFluctuations1987, Caputo-Sanvitto-TopologicalOrderThermal-2018}. To assure that our conclusions concerning the decays are meaningful notwithstanding the number of available parameters and statistical errors, we fit the data assuming stretching exponents $n_{d}$ fixed at 1.0 and 0.7 and evaluate the need for a value $n_{d} \neq 1$ based on the Akaike information criterion (AIC)~\cite{Akaike-NewLookStatistical-1974}.

Tracking the population dynamics as a function of pump frequency, we see asymmetric peaks in the amplitudes of the populations at the $n = 0$ and $n = 1$ minima (Fig.~\ref{fig:fig4}\textbf{E}), with no signal below the resonance, a maximum at the resonance and a plateau above it. The lifetimes (Fig.~\ref{fig:fig4}\textbf{F}) of the populations at the band minima show asymmetric enhancements, with peaks at the resonance pump frequency. Away from resonance, the $n = 1$ lifetime falls to values similar to those of the pumped magnons ($\sim$10~ns), while the $n = 0$ population lifetime also falls but not quite to the same extent. The AIC comparison (Fig.~\ref{fig:fig4}\textbf{D}) shows that the condensate decays near resonance require $n_{d} \neq 1$, i.e., the resonance condition produces decays of the type associated with other (quasi)ordered 2D systems such as polariton BECs and Ising magnets.

The mechanism for the creation of the condensate is responsible for the growth curves, which we plot in detail at resonance in Fig.~\ref{fig:fig4}\textbf{G}. The magnons directly generated by the microwaves appear immediately, while those at the bottom of the $n = 0$ and $n = 1$ bands appear much more slowly. Figure~\ref{fig:fig4}\textbf{H} shows the associated rise times as a function of the pump frequency; there is no minimum suggestive of particularly efficient production of magnon pairs involving both bands at resonance.

\section{Discussion}

The combination of threshold behavior in relation to the pump power, the narrow (resolution-limited) magnon peaks centered on the frequencies at the minima for the $n = 0$ and $n = 1$ modes, the nonexponential decay coupled with resonant lifetime enhancement, and the fact that the growth rate of the populations at the band minima is not accelerated near resonance but rather increases monotonically throughout the frequency range probed, lead us to conclude that 1) magnons are collected in condensate-like populations near the dispersion minima, 2) four-magnon scattering processes rather than direct microwave injection enhance this mechanism at the resonant pump frequency to distribute magnons to the minima, tipping the magnon density over a critical threshold and resulting in a longer-lived multiband condensate. 
In view of the nonequilibrium nature of the observed multiband condensate, the standard equilibrium variational argument by Nozieres~\cite{Griffin-Stringari-BoseEinsteinCondensation-1995} stating that a fractured condensate should be less energetically favorable, does not apply. The system is not free to minimize a free-energy functional at fixed chemical potential, and mode populations are instead determined dynamically by the competition between pumping, scattering, and decay.
Furthermore, the multiband nature makes the observed condensate profoundly different from those observed before, e.g. among magnons or polaritons.

Indeed, while pumping also allows for a coherent occupation of the two (idler and signal) modes in the optical parametric oscillator regime of polariton condensates~\cite{Christopoulos-Grandjean-RoomTemperaturePolaritonLasing-2007, Carusotto-Ciuti-QuantumFluidsLight-2012}, the modes are defined by single dispersions labeled by momentum $k$ rather than the multiple dispersions (due to thickness-induced quantization) discussed in our work. Furthermore, the effectively parallel pumping in our work precludes locking between the phases of the condensates and the pump, which is characteristic of polaritonic condensates. These differences delineate multiband magnon condensates observed in this work from the exciton-polariton condensates.

Having established the detailed nonequilibrium dynamics of a multicomponent magnon condensate as a function of pump amplitude for fixed pump frequency and as a function of pump frequency for fixed amplitude, it is natural to ask for a more general ``phase diagram", where the quotation marks imply that we are focusing only on signal intensities in a driven-dissipative system, as a function of both frequency and amplitude. Surveying our system using the external magnetic field to tune the magnon mode frequency relative to the pump ($\mathrm{\Delta}F_{P}$) and varying RF pump power (pump frequency fixed at 7.15 GHz) reveals a complex phase diagram (see Fig.~\ref{fig:fig4}\textbf{I}). For high fields ($\mathrm{\Delta}F_{P} < - 1\ GHz$), the magnon bands are above the pump half-frequency and no magnons can be excited. As the field is lowered, the half-frequency of the pump aligns directly with the $n = 0$ magnon band minima and a single component population is established at relatively low power, likewise, for the $n = 1$ band minima at the much lower field ($\mathrm{\Delta}F_{P} \approx 0.75\ GHz$). For most intermediate fields, magnons are injected into the $n = 0$ band away from its minima, however, they still populate the minima due to thermalization via magnon-magnon interactions. The formation of the condensate in this regime depends on the pump power: the condensate at $n=0$ is formed only for sufficiently high powers, see the boundary between the orange and the blue regions.
Finally, for a specific intermediate field ($\mathrm{\Delta}F_{P} \approx - 0.4\ GHz$), the relation of both bands and the pump frequency is such that magnons are excited into both bands simultaneously (the hatched region in Fig.~\ref{fig:fig4}\textbf{I}). This occurs at the resonance condition prescribed by the Feynman diagram of Fig.~\ref{fig:fig1}\textbf{C}. 
The formation of this multiband condensate also requires sufficient pump power, which is higher than for the $n=0$ condensate. Finally, at much lower magnetic fields, we populate the $n=1$ mode, see the green region. Overall, the phase diagram highlights three condensate regimes: (i) single-band $n=0$, (ii) multiband $n=0$ and $n=1$, and (iii) single-band $n=1$. All of them have power thresholds and require specific frequencies of the magnon modes relative to the pump.

\section{Summary}

We have found that thin films of high-quality YIG host a room temperature nonequilibrium multiband magnon condensate in both the $n = 0$ and $n = 1$ $k \parallel M_{0}$ magnon branches. The formation of this form of condensate is driven primarily by a resonant scattering process which agrees with the long-lived collective dynamics observed at the resonance. Thus, the physics in Fig.~\ref{fig:fig1}, namely a multiparticle scattering resonance leading to an optimized multicomponent condensate in a two-band system, appears validated. Given the widespread interest but limited experimental platforms and available theory, our results open a straightforward route to create, control and understand multiband condensates and their collective behavior, including normal modes and topological excitations. Particularly exciting opportunities are associated with the ease of imaging magnons in YIG using new techniques such as diamond NV sensors~\cite{Du-Yacoby-ControlLocalMeasurement-2017} and soft X-ray microscopy~\cite{Forster-Wintz-DirectObservationCoherent-2019, Forster-Schutz-NanoscaleXrayImaging-2019}. The resonance underpinning the multiband magnon condensate identified here could also be exploited in semiclassical and quantum devices where efficient generation and switching of magnons with sharply peaked momentum distributions is required.

\section{Materials and Methods}

\subsection*{Sample preparation}

The single-crystalline YIG films were grown on (111) GGG substrates by liquid-phase epitaxy (LPE) on 1-inch GGG wafers from a PbO-B$_2$O$_3$-based high-temperature solution using a standard dipping technique~\cite{Dubs-Wendler-LowDampingMicrostructural-2020, Dubs-Dellith-SubmicrometerYttriumIron-2017}. The 128 nm-thin YIG LPE film showed excellent spin wave damping with a full width at half maximum FMR linewidth of 1.0~Oe at 6.5~GHz and a Gilbert damping parameter $\alpha = 5\times10^{-5}$, giving FMR lifetimes of $\sim$350~ns. For RF excitation 2~$\mu\mbox{m}$ wide 200~nm thick Cu striplines were patterned on the samples by electron beam lithography and lift-off processing. To mitigate the otherwise significant charging during exposure, an aluminum layer was deposited on top of the resist. A micrograph of the completed devices is shown in Fig.~\ref{fig:fig1}\textbf{D} and \ref{fig:fig1}\textbf{E}.

\subsection*{Time-resolved Brillouin Light Scattering microspectroscopy}

We used a micro-BLS setup with $\sim$0.8~ns temporal and $\sim$300~nm spatial resolution to measure the magnon dynamics in our sample. The frequency of the incident light is shifted by inelastic scattering from magnons in the sample, thus the measurement of light intensity for a given frequency shift can be correlated with the density of magnons of said frequency. The experimental setup consists of a 473~nm continuous wave (CW) laser with an incident power of 0.85~mW focused through an objective lens with NA = 0.85. The laser is focused on the frontside (YIG film facing) to the side of the stripline. The reflected light is sent through a specialized Fabry-Perot interferometer and is detected with a single-photon counting detector. Quoted values for counts with noninteger values are due to averaging per unit time. The RF excitation is supplied by a signal generator (Anritsu) with high-speed switches, rise time $\sim$2~ns (Universal Microwave Components Corporation), controlling the pulse characteristics. For the phase diagram, a burst of 150~ns with a period 300~ns and a pump frequency of 7.15~GHz was used. For the detailed time-resolved measurements, a burst of 500~ns with a period of 2500~ns was used with a varying pump frequency and power. This latter measurement was explicitly configured to extract the temporal dynamics of the populations, i.e., giving us a maximum temporal resolution of 0.8~ns and `off period' for the RF $\sim$50$\times$ the observed decay times. All the quoted powers are instantaneous powers onto the sample. The spectrometer, photon detector and switches are all locked to a common clock such that photon detection events are then binned both in frequency shift and time. The sample is magnetized in-plane using a permanent magnet and oriented such that the micropatterned transmission line is perpendicular to the external field, in the so-called parallel pumping configuration.

\subsection*{Data analysis}

For the measurements of magnon dynamics for varying pump power and pump frequency, the different magnon populations were distinguished by separate frequency windows. These windows were determined via a fit of the sum of 3 normal distributions (for BLS, the Gaussian smoothing due to the poor frequency resolution of the instrument dominates all other line shapes) with a constant background. The center of each of these 3 normal distributions was taken to be the central frequency of each population. The frequency window around the central frequency was defined as the mean full width at half maximum determined by the fits and common for all populations. In particular, for the 50~mT sweeps shown in Figs.~\ref{fig:fig2}--\ref{fig:fig4}, the frequency windows for the zeroth order populations, pumped magnons, and first order populations, with pump frequency 6.7~GHz were found to be $2.59$, $3.35$, and $4.12 \pm 0.13$~GHz. There was found to be a small frequency offset (0.15~GHz) between the pumped magnon populations and the set pump frequency, which was due to the calibration of the spectrometer. Therefore, all reported frequencies are shifted to take this into account.

\section*{Author contributions}
J.B., P.S., J.R., D.G., A.B., and G.A. designed research; J.B., P.S., and K.B. performed research; S.F., S.W., and C.D. contributed new reagents/analytic tools; J.B., P.S., and A.B. analyzed data; and J.B., P.S., and G.A. wrote the paper.

\section*{Competing interests}
The authors declare no competing interest.

\section*{Data availability}
All study data are included in the article and/or SI Appendix.

\section*{Acknowledgments}
We are especially grateful to the late Peter Littlewood not only for his very insightful and helpful comments as a referee of this manuscript, but also for the guidance, support and friendship which two coauthors (AB and GA) have benefitted tremendously from over many decades. Giancarlo Corradini (EPFL) is thanked for his assistance with the wire bonding the samples and Oleksii Surzhenko (INNOVENT) for the ferromagnetic resonance (FMR) line width measurements. Financial support came from the European Union's Horizon 2020 Research and 
Innovation Programme Marie Sk{\l}odowska-Curie Grant Agreement No. 66566 (J.B.), the European Research Council under the European Union's Seventh Framework Program Synergy HERO SYG-18 810451 (J.B., P.S., A.B., and G.A.) Knut and Alice Wallenberg Foundation KAW 2018.0104 (P.S. and A.B.), the Swiss NSF 16301 (K.B.) and the Deutsche Forschungsgemeinschaft-271741898 (C.D.)



\bibliography{library-mBEC}

\end{document}